\documentclass[preprint,12pt]{elsarticle}

\usepackage{amssymb}
\usepackage{amsmath}

\usepackage{fullpage}
\usepackage{epsfig}
\usepackage{comment}
\usepackage{xcolor}
\usepackage{longtable}
\usepackage{multirow,array}

\usepackage{graphicx}
\usepackage{caption}
\usepackage{subcaption}

\usepackage{enumitem}
\usepackage{adjustbox}
\usepackage{tcolorbox}
\usepackage{listings}
\usepackage{matlab-prettifier}
\lstdefinestyle{PythonStyle}{
    language=Python,
    basicstyle=\ttfamily\footnotesize, 
    keywordstyle=\color{blue}\bfseries,
    identifierstyle=\color{black},
    stringstyle=\color{orange},
    commentstyle=\color[rgb]{0.13,0.54,0.13}, 
    showstringspaces=false,
    columns=fullflexible, 
    breaklines=true,
    frame=single, 
    frameround=tttt,
    rulesepcolor=\color{gray},
    numbers=left, 
    numberstyle=\tiny\color{gray},
    stepnumber=1,
    tabsize=4,
    captionpos=b, 
    backgroundcolor=\color{gray!10}, 
    literate={å}{{\aa}}1
             {ø}{{\o}}1    
}

\usepackage[utf8]{inputenc}

\usepackage{color}
\usepackage{xspace}
\usepackage{cancel}

\usepackage{parskip}
\usepackage{hyperref}

\journal{Cyber-Physical Energy Systems}

\begin{document}

\begin{frontmatter}



\title{Cenergy3: An Open Software Package for \\City Energy 3D Modeling} 

\author{Shiliang Zhang}
\author{Sabita Maharjan}
\affiliation{organization={University of Oslo},
             city={Oslo},
             postcode={0373},
             country={Norway}}



\begin{abstract}
The efficient management and planning of urban energy systems require integrated three-dimensional (3D) models that accurately represent both consumption nodes and distribution networks. This paper introduces our developed approach and openly released software that automate the generation of digital 3D urban energy model from open data. We synthesize data from OpenTopography, OpenStreetMap, and Overture Maps in generating 3D models. The rendered model visualizes and contextualizes distribution power grids alongside the built environment and transportation networks. Our developed software, including an open python library and a free API, provides interactive figures for the 3D models. The rendered models are essential for analyzing infrastructure alignment and spatially linking energy demand nodes (buildings) with energy supply (utility grids). The developed API leverages standard \texttt{Web Mercator} coordinates (\texttt{EPSG:3857}) and \texttt{JSON} serialization to ensure interoperability within smart city and energy simulation platforms. We also provide a graphic user interface (GUI) where end-users can access our API via a cloud-based server, regardless of their programming skills and what devices and platforms their are using. We anticipate that our approach and software can support field researchers, developers, end-users, and policy-makers in a varieties of applications like urban energy monitoring, demand-supply analysis, and energy digital twins.
\end{abstract}

\begin{graphicalabstract}
\includegraphics[width=0.98\textwidth]{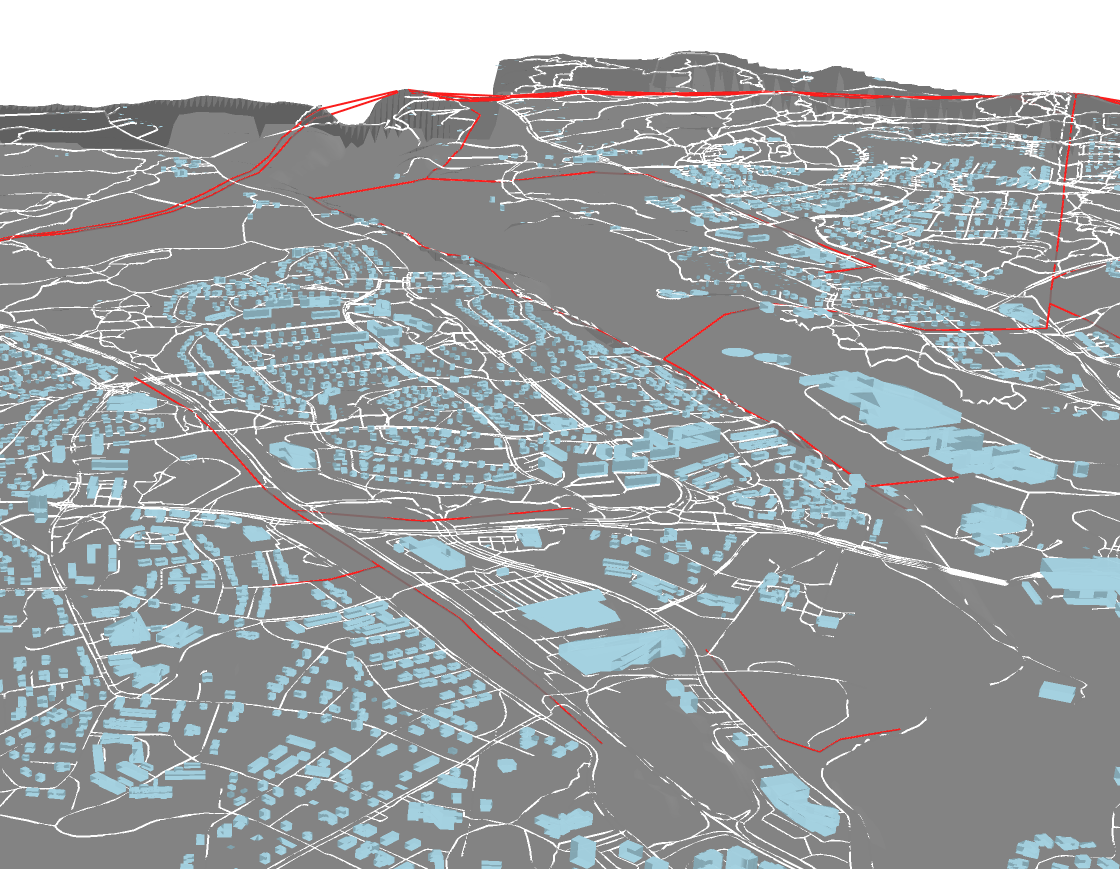}
\end{graphicalabstract}

\begin{highlights}
\item A novel approach to retrieve, compile, and visualize the power grid and the contextual environments in urban areas from open source data.
\item A released open Python library to automate 3D modeling of power grid, terrain, and buildings for any area of interest.
\item A free application programming interface (API) that provides a cloud-computing based service for 3D city energy modeling.
\item A browser-based graphic user interface (GUI) for accessing our API, regardless of end-users' programming skills or what devices they are using.
\end{highlights}


\end{frontmatter}




\section{Introduction}

The modern urban landscape is currently experiencing a digital shift, marked by a rapid accumulation of geospatial datasets pertinent to energy systems~\cite{ujang2025introduction}. This data covers a broad spectrum of variables, such as consumption metrics, energy generation, energy price, and physical infrastructure configurations, \textit{etc}~\cite{hersbach2019era5,weinand2019spatial}. Analyzing these datasets is vital for extracting insights and developing knowledge necessary to achieve international sustainability targets~\cite{mallala2025forecasting}. In particular, geospatial methodologies facilitate the planning of energy infrastructure, power grid resilience during renewable energy integration, and power flow optimization~\cite{andres2025hot,ahshan2025geospatial}. Moreover, such analysis enables demand-supply assessments, encompassing predictive modeling and long-term urban development forecasting~\cite{10538433,oskar2025exploring}.

However, utilizing high-resolution energy information, such as granular smart meter data, raises significant privacy and ethical concerns~\cite{zhang2025data}. Although these datasets offer comprehensive details regarding energy usage, they are inherently sensitive~\cite{cheikh2026energy}. Specifically, high-frequency energy data may inadvertently disclose household occupancy, lifestyle preferences, and daily habits, thereby compromising individual privacy~\cite{wylde2022cybersecurity}. This vulnerability frequently causes utility providers and scholars to be reluctant to share data, which subsequently limits innovation in the field of urban energy management.

Open-source geospatial data serves as a privacy-compliant alternative for extensive energy modeling and urban planning~\cite{SALVALAI2024114500,el2024comprehensive}. Platforms such as OpenStreetMap~\cite{9119753} offer high-fidelity, anonymized proxies for energy-relevant features. These include building geometries, transportation accessibility, and the spatial layout of physical assets~\cite{moroz2026urban,khatiwada2026urban}, all accessible without compromising sensitive personal information. By leveraging these open resources, researchers can perform rigorous analyses while upholding data ethics and ensuring anonymity. 

Nevertheless, a major challenge persists in converting fragmented, data-intensive analyses into intuitive insights, visual~\cite{quinones2025data}. Despite the wealth of raw data, there is a distinct lack of robust and open tools capable of visualizing energy infrastructures within the intricate spatial context of the urban fabric~\cite{11364211}. Conventional static reports and 2D mapping techniques often fail to represent the interplay between topography, urban density, and the physical routing of power distribution networks~\cite{yan2023multi,kapp2023predicting}. This lack of contextual visualization can create communications gaps between urban planners, engineers, and policymakers, who may interpret identical data differently in the absence of a unified spatial reference.

To address this deficiency, this paper introduces an open-source software suite designed to convert public data into immersive, geospatially-aware 3D urban environments. This package comprises a Python library and an API hosted on a cloud server, both of which are provided free of charge to the public. The software automates the generation of 3D models by integrating terrain data from OpenTopography~\cite{krishnan2011opentopography}, detailed building semantics from Overture Maps~\cite{shiell2025overture}, and infrastructure topology (roads and power grids) from OpenStreetMap. The Graphical Abstract illustrates an representative 3D model generated by the tool. By situating power grid infrastructure within the built environment and transportation networks, this software provides a critical tool for spatially correlating energy data with physical city structures, presenting a novel advancement for urban energy research and planning. The contributions of this study are summarized as follows.

\begin{itemize}
    \item We develop an open-data-driven approach that empowers simultaneous visualization of power grids, built environments, and transportation networks that facilitates a holistic understanding of the urban energy nexus.
    \item We develop an open software package, \texttt{Cenergy3}, comprising a Python library and a cloud-based API, which automates the synthesis of terrain data, building semantics, road networks, and energy infrastructure topology.
    \item We demonstrate the utility of the released library and API, and provide programming examples for direct use of our package. We also develop a browser-based graphic user interface (GUI) so that end-users without code expertise can access and utilize our package conveniently. This also lowers the barrier to the entry of our package for stakeholders in resource-constrained environments.
\end{itemize}

The rest of this paper is organized as follows. Section~\ref{sec:method} presents the method we developed for the data synthesis from open sources and the visualization of 3D models. Section~\ref{sec:software} introduces the developed Python library and the cloud-based API. Section~\ref{sec:demo} demonstrates the usage of our library in Python and MATLAB programming examples, and the usage of our API via our developed browser-based GUI. Section~\ref{sec:interoperability} clarifies the interoperability of our library and API, and we conclude this paper in Section~\ref{sec:conclusion}.

\section{The methodology behind the developed software package}\label{sec:method}

The developed package \texttt{Cenergy3} is constructed entirely upon public, open-access data repositories. This package enables client-side applications to generate 3D models via a dedicated Python library or through standard HTTP request to our hosted API. The computational pipeline of the software is structured into four phases: Geocoding, terrain processing, vector data draping, and building extrusion, as detailed in the following.

\begin{figure}[tbhp]
    \centering
    \includegraphics[width=0.4\linewidth]{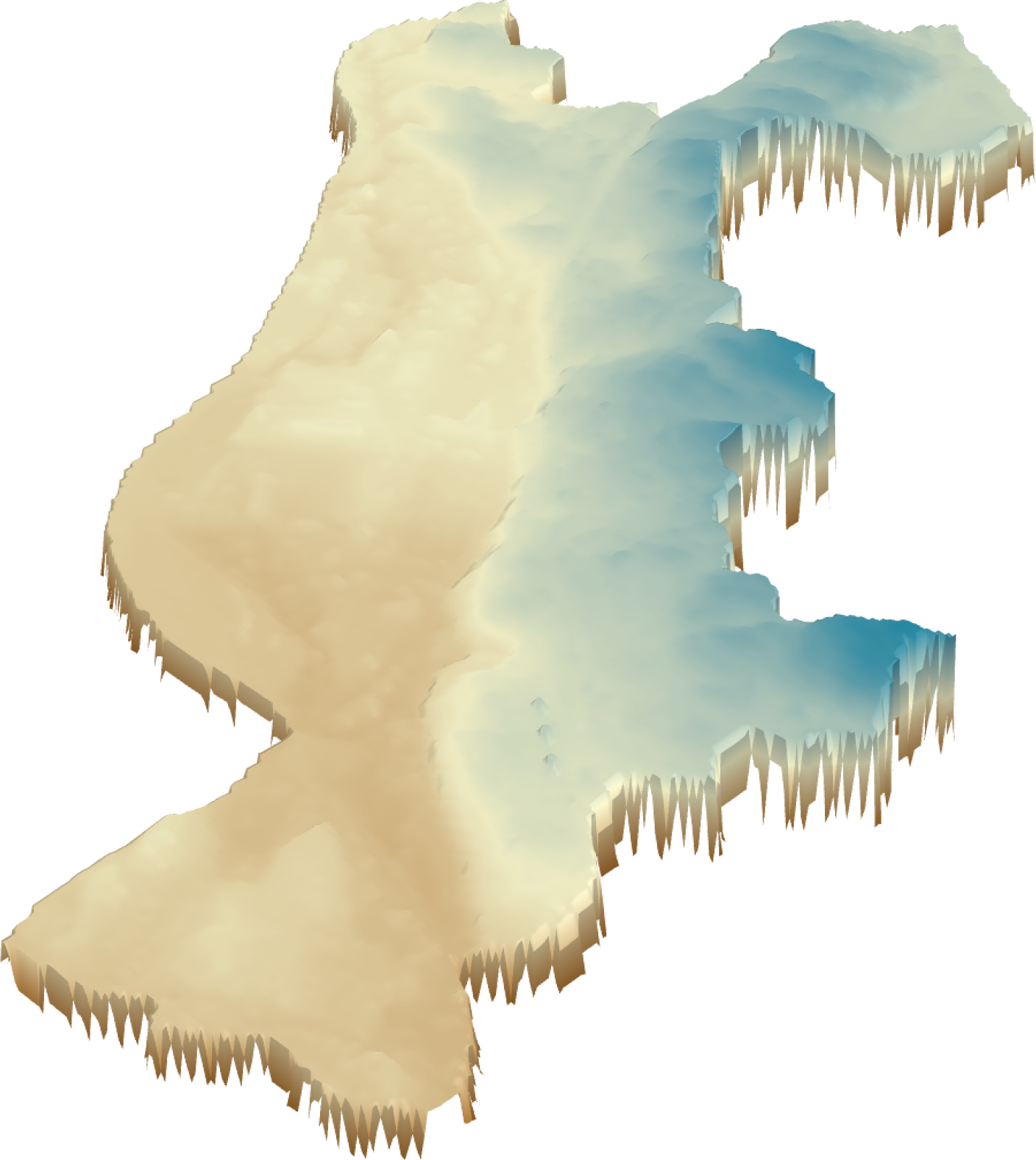}
    \caption{An example of the visualized terrain for the area of Alna, Oslo, Norway.}
    \label{fig:terrain}
\end{figure}

\subsection{Geocoding}
Upon receiving a user request, \texttt{Cenergy3} employs the \texttt{OSMnx}~\cite{boeing2025modeling} toolkit to translate the target place name into a georeferenced polygon boundary sourced from OpenStreetMap~\cite{bennett2010openstreetmap}. From this polygon, a precise bounding box is derived to delineate the spatial extent of the study area. This coordinate-based bounding box serves as the primary reference for subsequent data acquisition, ensuring rigorous spatial synchronization across all layers.

\begin{figure}[tbhp]
    \centering
    \includegraphics[width=0.65\linewidth]{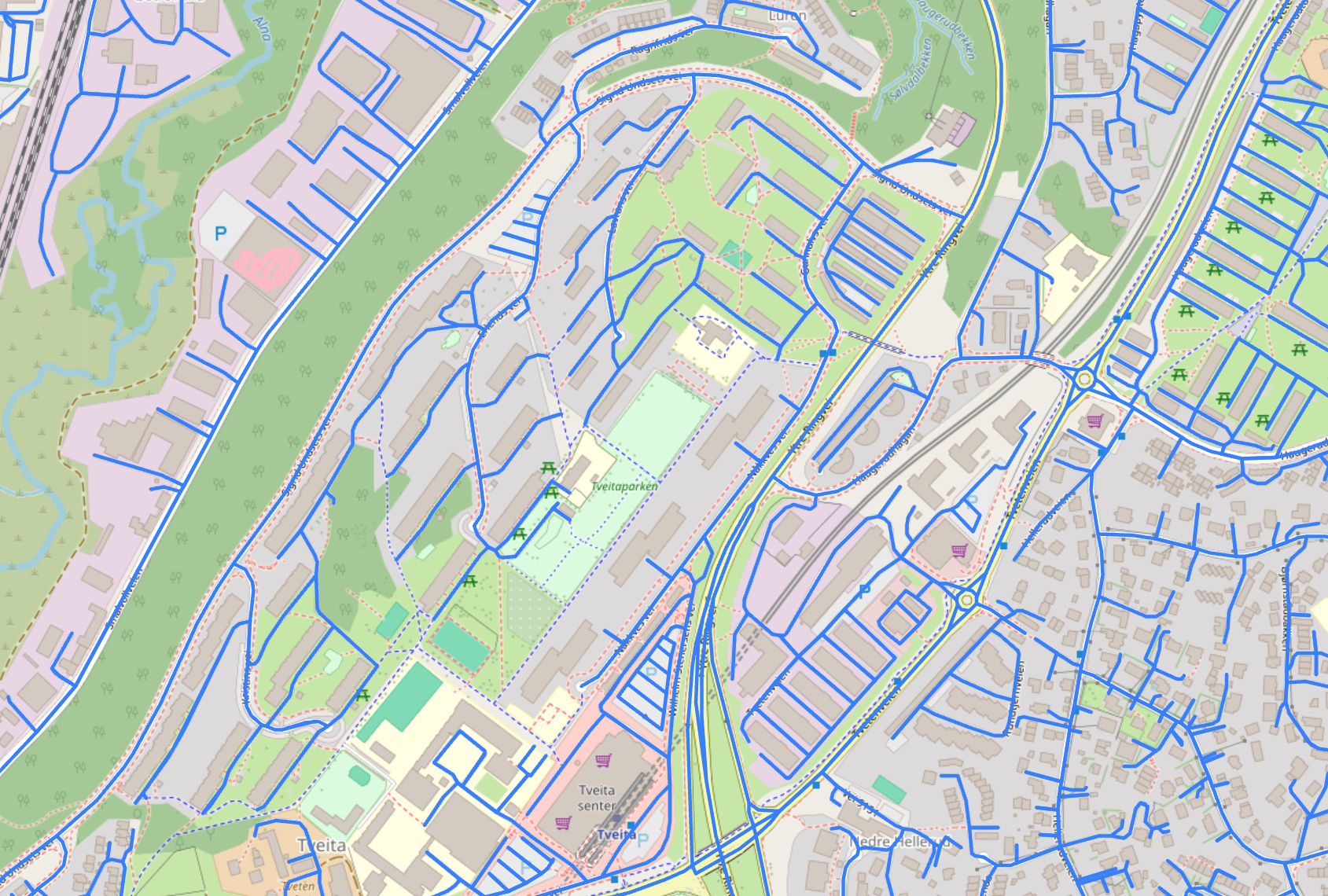}
    \caption{Retrieved road networks from OSM within the area of Alna, Oslo, Norway. The blue lines indicate the road networks in the city.}
    \label{fig:roads}
\end{figure}

\subsection{Terrain processing}
The digital elevation model (DEM)~\cite{mukherjee2013evaluation} serves as the structural foundation for the 3D environmental reconstruction. We retrieve elevation data from the global copernicus 30m (COP30) DEM via the OpenTopography interface. Accessing this service requires a API key, which is available and free from the OpenTopography platform. To maintain metric precision during 3D visualization, we reproject the COP30 DEM data from its original coordinating reference system~\cite{kumar2023referencing} (EPSG:4326) to the Web Mercator projection~\cite{stefanakis2017web} (EPSG:3857). 

The reprojected elevation data is subsequently transformed into a triangular mesh to represent the local topography. We present an illustrative example of processed terrain in Fig.~\ref{fig:terrain}. Specifically, the algorithm generates geometric vertices for every valid pixel within the COP30 dataset. The terrain faces are then constructed by triangulating the centers of adjacent pixels, thereby standardizing the elevation geometry prior to final rendering.

\subsection{Vector data draping}
Integrate 2D vector information, such as road networks and power distribution lines, with a 3D topographic surface presents a significant challenge regarding vertical alignment. To solve this, the \texttt{Cenergy3} package utilizes a specialized draping mechanism.

\begin{figure}[tbhp]
    \centering
    \includegraphics[width=0.5\linewidth]{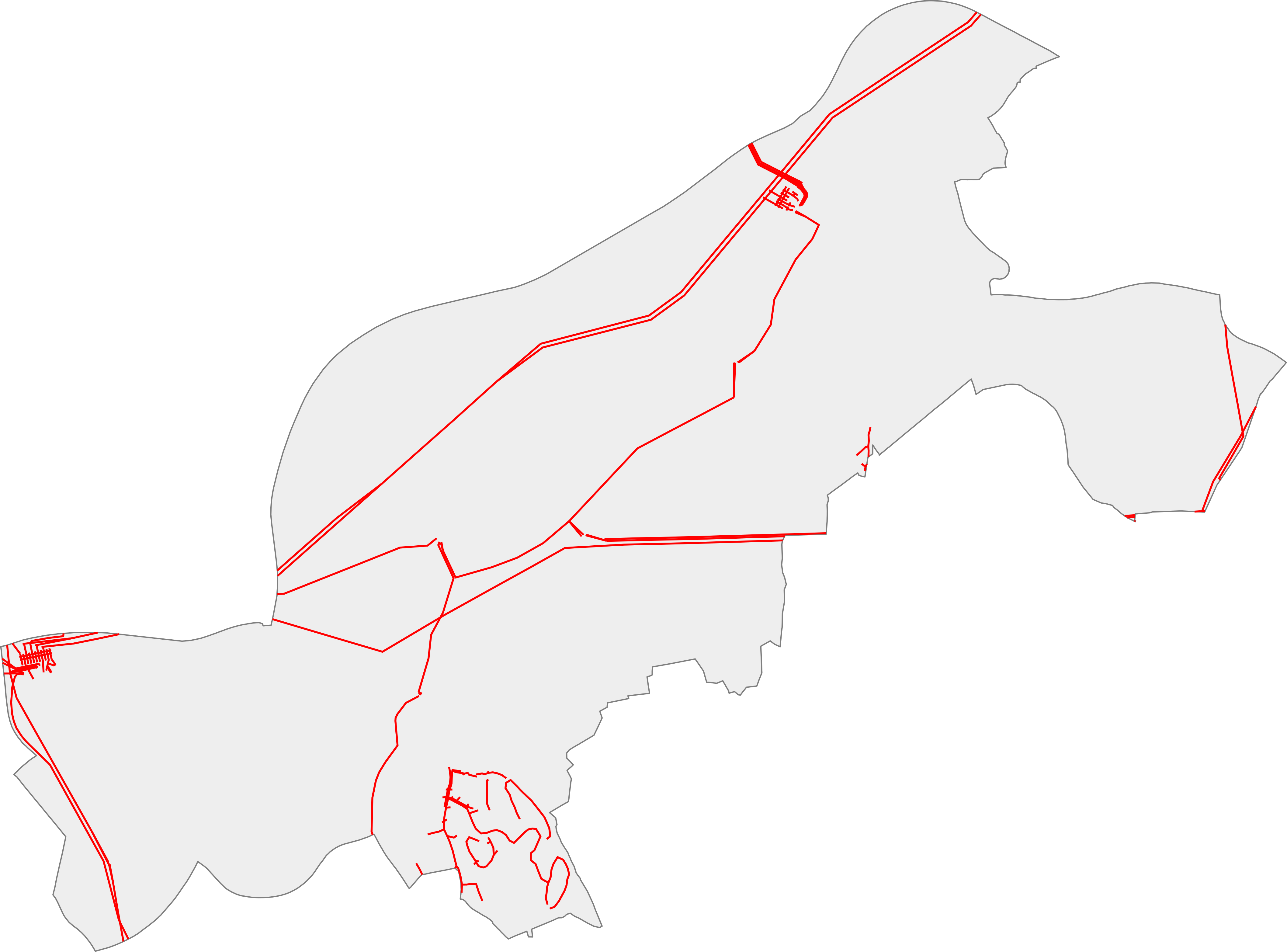}
    \caption{Retrieved power lines from OSM for the area of Alna, Oslo, Norway. The red lines indicate power cables in the city.}
    \label{fig:powerlines}
\end{figure}

The process begins by extracting road network geometries for the target area from OpenStreetMap (OSM), an example of which is shown in Fig.~\ref{fig:roads}. The road edges are then discretized, with elevation values interpolated for each point from the underlying COP30 DEM. Consequently, roads are rendered as 3D-aware paths that follow the terrain's contours. Our software applies the same workflow to power line vectors retrieved from OSM (see Fig.~\ref{fig:powerlines}). By draping these vectors over the digital terrain, our software effectively visualize energy corridors in direct relation to the physical landscape.

\begin{figure}[tbhp]
    \centering
    \includegraphics[width=0.65\linewidth]{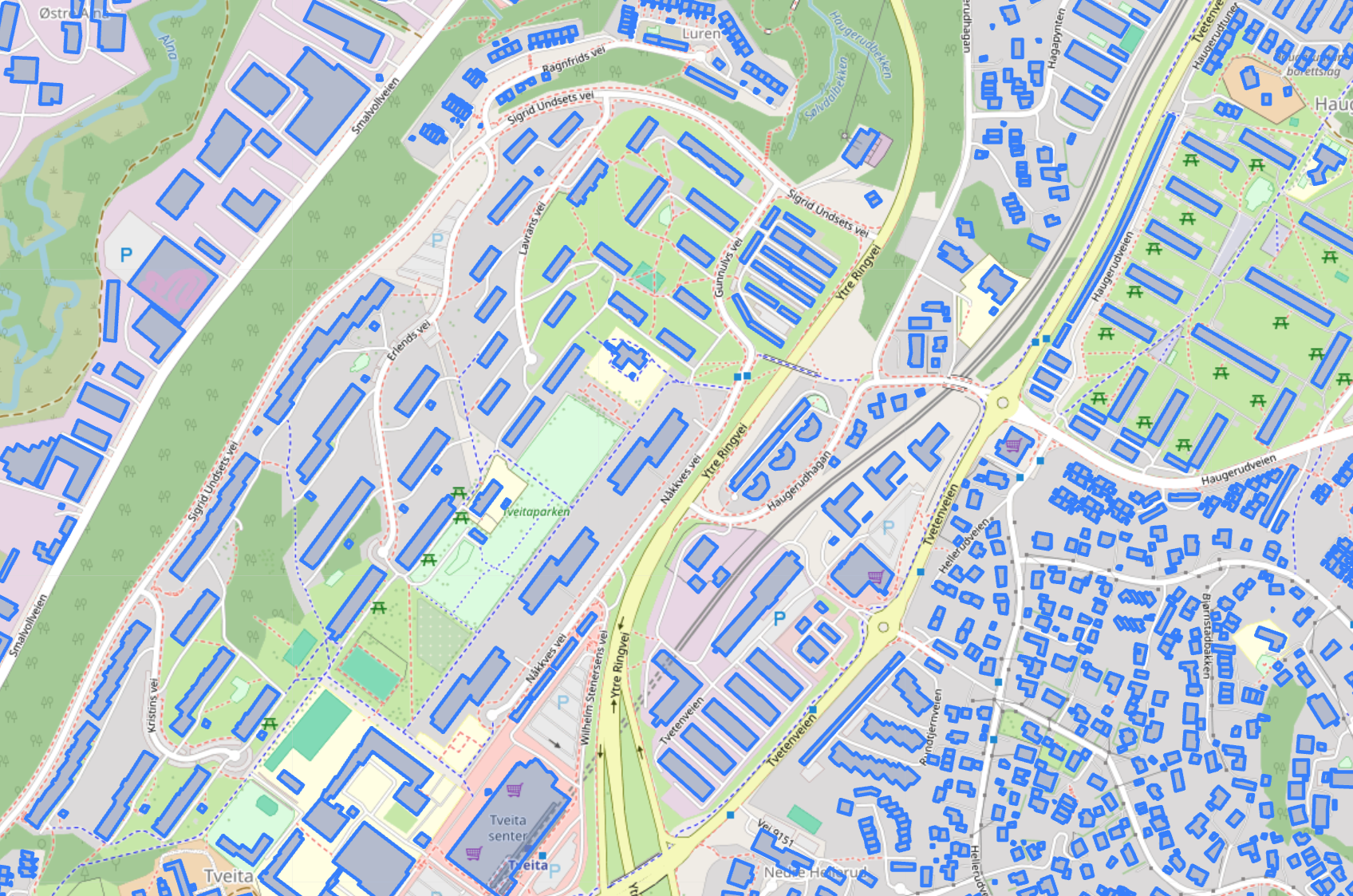}
    \caption{Retrieved building footprints from OSM within the area of Alna, Oslo, Norway. The polygons in blue lines indicate buildings in the city.}
    \label{fig:buildings}
\end{figure}

\subsection{Building extrusion}
A common limitation in OSM datasets is the absence of comprehensive building height information. \texttt{Cenergy3} addresses this by incorporating high-fidelity data from \texttt{Overture Maps}\footnote{\url{https://overturemaps.org}, accessed 2026-03-04.}. The procedure begins with the extraction of building footprints from OSM, as shown in Fig.~\ref{fig:buildings}. These footprints are then cross-referenced with Overture Maps to retrieve corresponding height attributes.

\begin{figure}[tbhp]
    \centering
    \includegraphics[width=0.65\linewidth]{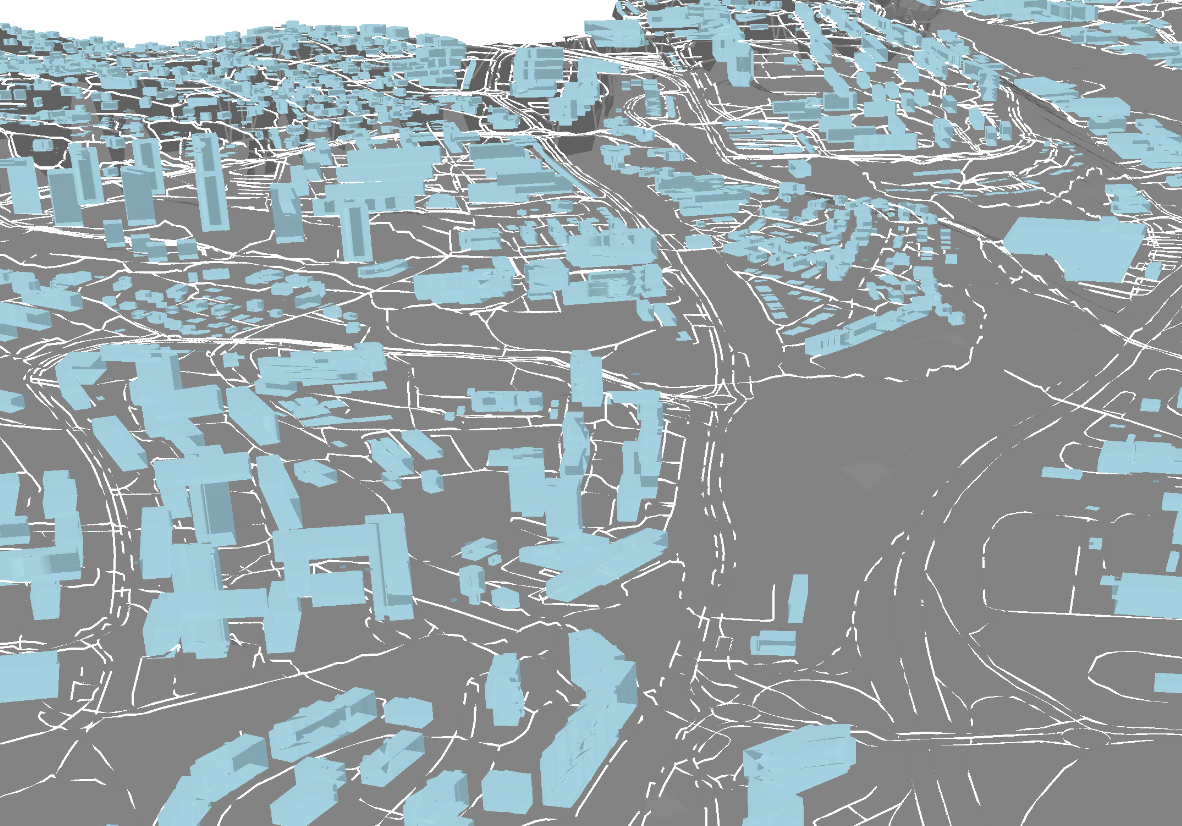}
    \caption{An example of extruded buildings within the area of Alna, Oslo, Norway.}
    \label{fig:building_3d}
\end{figure}

For each building, our software generates a base polygon at the local terrain elevation and corresponding roof polygon adjusted by the specific building height. We then synthesize the vertical surfaces of the building by connecting the base and roof polygon horizontal planes, resulting in a complete 3D mesh for every building. A representative example of these extruded urban buildings is shown in Fig.~\ref{fig:building_3d}.

\section{Cenergy3: openly released Python library, free API, and publicly accessible GUI}\label{sec:software}

\subsection{Python library}

We have openly released our code for the visualization of city energy 3D models. We develop the code in Python, and we release the code as an open Python library via PyPI.org. The code underpinning the released library can be accessed via our Github webpage\footnote{\url{https://github.com/slzhang-git/cenergy/}, accessed 2026-03-11.}, and the released library is documented on PyPI.org\footnote{\url{https://pypi.org/project/cenergy/}, accessed 2026-03-11}. We open-source our code so that developers and researchers can build more upon our efforts in customized developments and studies.

\subsection{Application programming interface (API)}

We have developed an cloud-based API based on our Python library. This effort aims to release the computational limits of end-users' local environments. Furthermore, an API is also the prerequisite for the development of a graphic user interface (GUI) that fully liberate the end-users from their programming skills and their dependencies of local environments and devices.

We deploy our API on a cloud server provided by DigitalOcean\footnote{\url{https://www.digitalocean.com}, accessed 2026-03-11.}. We use FastAPI~\cite{lathkar2023high} to build and deploy a RESTful API~\cite{biehl2016restful} on the cloud server to provide standard web app service. We dedicate 4G RAM\footnote{This RAM might change over the time depending on our resources. We use 4G RAM from 2026-01.} and one CPU to our deployed API, which should be enough for the 3D model generation at district levels in cities.

\subsection{Graphic user interface (GUI)}

We develop the GUI for our API so that end-users can access 3D city energy models regardless of their programming skills and constraints of their local computation environments. With our GUI, an end-user is free from installing any software and coding efforts. 

\begin{figure}
    \centering
    \setlength{\fboxsep}{0pt}  
    \setlength{\fboxrule}{1pt} 
    \fbox{\includegraphics[width=0.5\linewidth]{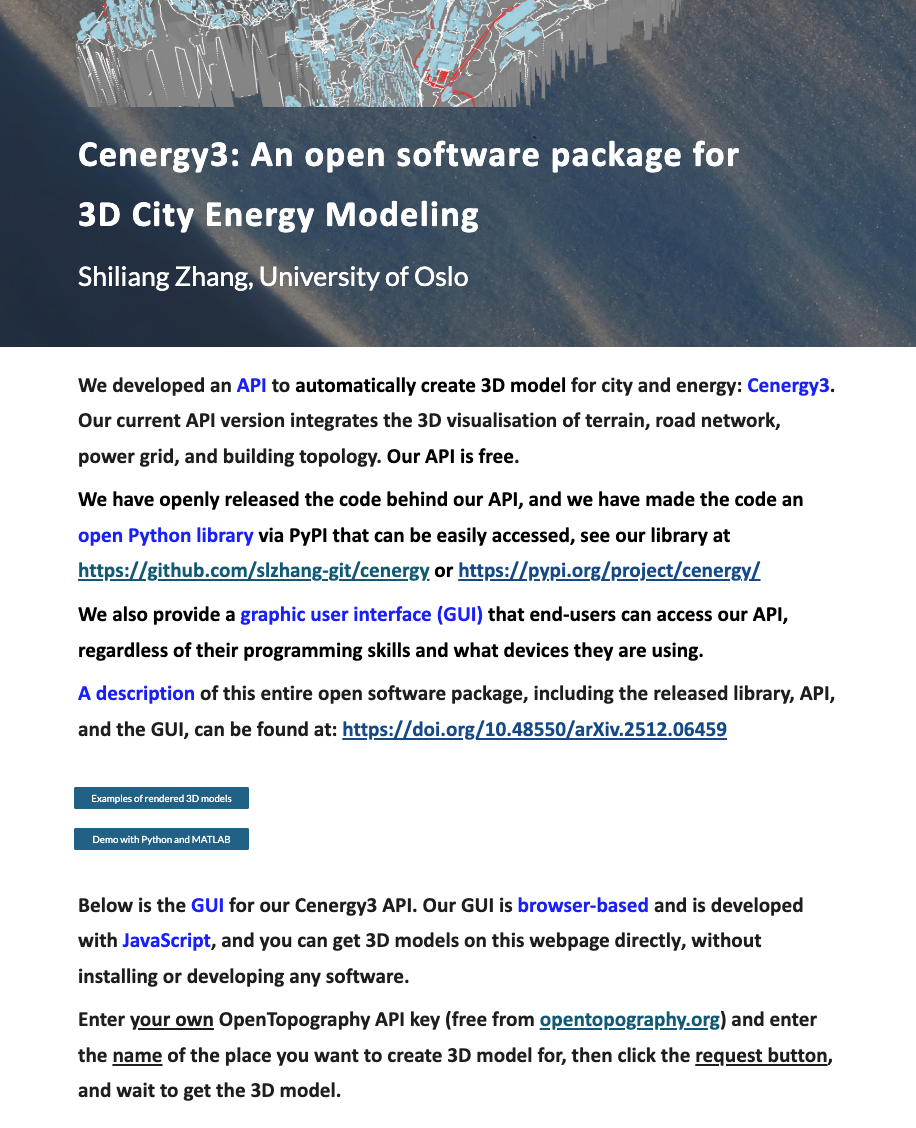}}
    \caption{The browser-based GUI we developed for accessing our API. It is a browser-based GUI that can be publicly accessed at \url{https://sites.google.com/view/cenergy3/home}}
    \label{fig:interface}
\end{figure}

Our GUI is browser-based and is developed with JavaScript, and we deploy our GUI via Google Sites\footnote{\url{https://sites.google.com/}, accessed 2026-03-11.}. Fig.~\ref{fig:interface} shows the deployed GUI, where end-users only need to feed in a free API key and the name of the place they want to request for the 3D models. We also provide some demos for the requested models in the GUI webpage.

\begin{figure}[tbhp] 
    \centering\caption{A programming example of accessing and using our released Python library}\label{lst:example_library}
    \begin{lstlisting}[style=PythonStyle]
!pip install cenergy
from cenergy3 import generate_3d_model, plot_3d_model, save_3d_model
api_key = "123456789123456789123456789" # Change to your own OpenTopography API key
target_place = "Rousay-Orkney Islands-Scotland" # You can change to the name of the place
fig_json = generate_3d_model(api_key=api_key, target_place=target_place)
plot_3d_model(fig_json) # Visualize the generated 3D model
save_3d_model(fig_json) # Save the model to your local disk
    \end{lstlisting}
\end{figure}

\section{Demonstration of how to use our Python library and our API}\label{sec:demo}

\subsection{Demo of our Python library}

The usage of our released Python library is intuitive, and we provide a programming example below in a Google Colab environment:

\begin{figure}[tbhp] 
    \centering\caption{A Python example to construct request to our API, for the place Rousay, Orkney Islands, Scotland.}\label{lst1}
    \begin{lstlisting}[style=PythonStyle]
import requests
import json
from io import StringIO
api_key = '111222333444555666777888999'   
# Replace it by your OpenTopography API key
target_place = 'Rousay-Orkney Islands-Scotland' 
# Please replace the name in your case
# Below we show how to construct the url for the request to our API
BASE_URL = "https://cenergy3-qjbps.ondigitalocean.app"
api_request_url = f"{BASE_URL}/{api_key}/{target_place}"
# Fetch data from API by sending a request
try:
    response = requests.get(api_request_url)
    response.raise_for_status() # Check for the status of the request
    figure_dict = response.json()   # Gain the requested data in json format
except requests.exceptions.RequestException as e:
    print(f"Error fetching data: {e}")
    \end{lstlisting}
\end{figure}

In the code in Fig.~\ref{lst:example_library}, we input the \textit{OpenTopography} API key and the name of the target area, and then generate, visualize, and save the 3D model for the target area. Note that this additional API key is free and can be obtained from the \textit{OpenTopography} platform\footnote{\url{https://opentopography.org}, accessed November 27, 2025.}. The target area utilizes a specific, URL-friendly format (\textit{e.g.}, \texttt{Rousay-Orkney Islands-Scotland}) to define the precise area of interest.

Through the code in Fig.~\ref{lst:example_library}, you install our library to your local environment. Then all the computations, including the data gathering and integrating from open sources, the formatting and visualizing of the data, and post-processing and saving the visualized results, will be conducted using the computing resources of your local environment. In this way, you avoid any latency that otherwise occurs when accessing 3D models from our cloud server via our released API, as detailed in Section~\ref{sec:demo_api}. Note that the response time depends on your local computation power when using our library on your local environment.

\subsection{Demo of accessing our API with Python and MATLAB programming}\label{sec:demo_api}

\begin{figure}
\centering\caption{An MATLAB example of constructing request to our API.}\label{lst:api_code_matlab}
\begin{adjustbox}{width=0.8\textwidth}
\begin{tcolorbox}[colback=gray!5!white,colframe=gray!75!black]
    \begin{lstlisting}[basicstyle=\tiny, frame=single, language=MATLAB, style=Matlab-editor]
api_key = '111222333444555666777888999'; % Replace it
target_place = 'Avalon, Los Angeles County, United States';

BASE_URL = "https://cenergy3-qjbps.ondigitalocean.app"; 
api_endpoint_url = sprintf('%s/%s/%s', BASE_URL, api_key, target_place);
disp(['Fetching data from: ' api_endpoint_url]);

% Fetch data from API 
try
    options = weboptions('RequestMethod', 'get', 'ContentType', 'text', 'Timeout', 3000);
    response_text = webread(api_endpoint_url, options);
    figure_struct = jsondecode(response_text);
    disp(' ');
    disp('Successfully fetched and decoded the JSON structure.');
catch ME % Catch errors
    disp(' ');
    disp(['Error during data fetching or processing: ' ME.message]);
end
\end{lstlisting}
\end{tcolorbox}
\end{adjustbox}
\end{figure}

This section shows how to use our API via standard HTTP requests, in both Python and MATLAB examples. The client-side interaction involves two steps: constructing the request URL and fetching the data, followed by rendering the resulting 3D model.

The API accepts parameters directly within the URL path. As shown in Fig.~\ref{lst1} in Python and Fig.~\ref{lst:api_code_matlab} in MATLAB, the request is formulated by concatenating the base URL with the necessary parameters: the \textit{OpenTopography} API key and the name of the target area. The system relies on the standard Python \texttt{requests} library to manage the \texttt{HTTP GET} transaction. The response body contains the 3D model, serialized as a complete \texttt{JSON} object describing the 3D representation.


The final step is to convert the retrieved \texttt{JSON} back into an interactive visualization object. As detailed in Fig.~\ref{lst:api_code_visualization_python} in Python and Fig.~\ref{lst:api_code_visualization_matlab} in MATLAB, the client-side code utilizes different ways to deserialize the \texttt{JSON} object. Particularly, we in the Python example use \texttt{plotly.graph\_objects.Figure} to deserialize the \texttt{JSON}, while we use \texttt{Ploty} and save the visualized results in \texttt{HTML} file in our MATLAB example. This deserialization process instantly reconstructs the 3D scene. The use of \texttt{Plotly} guarantees that the output is immediately interactive, allowing the user to rotate, zoom, and inspect the contextual relationships between the energy infrastructure and the environment.

\begin{figure}[tbhp] 
    \centering\caption{A Python example of visualizing the requested data from our API}\label{lst:api_code_visualization_python}
    \begin{lstlisting}[style=PythonStyle]
import plotly.graph_objects as go
try:
    fig = go.Figure(figure_dict)
    fig.show()  # Display the interactive figure for the 3D model
except Exception as e:
    print(f"\nError in displaying the requested data: {e}")
    \end{lstlisting}
\end{figure}

\begin{figure}[tbhp] 
    \centering
    \includegraphics[width=0.65\linewidth]{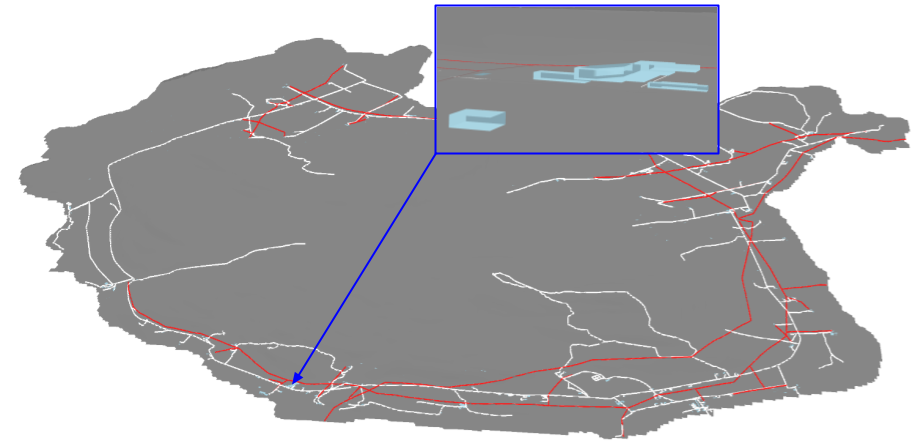}
    \caption{Visualized 3D model for the place of Rousay, Orkney Islands, Scotland. According to the log of our API, for this area, we collect 129,652 records of elevations, 988 road segments, 36 power lines, and 716 buildings with height. All the collected data is from open sources.}
    \label{fig:result_api_python}
\end{figure}

Fig.~\ref{fig:result_api_python} presents the visualization of 3D model for the place of Rousay, Orkney Islands, Scotland, when using Python programming to request our API. From this figure, we can observe that this place is a flat area, with buildings, roads, and power lines distributed around its boundary. 

\begin{figure}
\centering\caption{An MATLAB example of visualizing the requested data from our API}\label{lst:api_code_visualization_matlab}
\begin{adjustbox}{width=0.73\textwidth}
\begin{tcolorbox}[colback=gray!5!white,colframe=gray!75!black]
\begin{lstlisting}[basicstyle=\tiny, frame=single, language=MATLAB, style=Matlab-editor]
% Read JSON file in a browser
outHtml = fullfile('3D_visualization.html');
htmlTemplate = [...
'<!doctype html>\n' ...
'<html>\n' ...
'<head>\n' ...
'  <meta charset="utf-8">\n' ...
'  <title>Plotly JSON Viewer</title>\n' ...
'  <!-- Load Plotly from CDN -->\n' ...
'  <script src="https://cdn.plot.ly/plotly-2.29.1.min.js"></script>\n' ...
'  <style>body{margin:0;font-family:Arial,Helvetica,sans-serif} #plot{width:100vw;height:100vh;}</style>\n' ...
'</head>\n' ...
'<body>\n' ...
'  <div id="plot"></div>\n' ...
'  <script>\n' ...
'    // The JSON figure object is inserted below. It should be an object with "data" and optionally "layout" and "frames".\n' ...
'    var fig = '];
htmlTail = [...
';\n' ...
'    // If fig is an array of traces, convert to {data: fig}\n' ...
'    if (Array.isArray(fig)) {\n' ...
'      Plotly.newPlot("plot", fig, {});\n' ...
'    } else if (fig && fig.data) {\n' ...
'      var layout = fig.layout || {};\n' ...
'      var config = fig.config || {};\n' ...
'      Plotly.newPlot("plot", fig.data, layout, config);\n' ...
'    } else {\n' ...
'      // If the structure is unexpected, try to print it and attempt to plot\n' ...
'      console.warn("Loaded JSON not recognized as Plotly figure. Attempting to plot if possible.");\n' ...
'      try { Plotly.newPlot("plot", fig, {}); } catch(e){ document.getElementById("plot").innerText = "Unable to render JSON as Plotly figure. See console for details."; console.error(e); }\n' ...
'    }\n' ...
'  </script>\n' ...
'</body>\n' ...
'</html>\n'];
% Combine and write HTML
fid = fopen(outHtml, 'w', 'n', 'UTF-8');
if fid == -1, error('Cannot create output HTML file: %s', outHtml); end
fprintf(fid, htmlTemplate); 
fprintf(fid, '%s', figure_struct); 
fprintf(fid, htmlTail);  
fclose(fid);
% Open in system browser
fprintf('Wrote HTML to: %s\n', outHtml);
web(outHtml, '-browser'); 
\end{lstlisting}
\end{tcolorbox}
\end{adjustbox}
\end{figure}

\begin{figure}
    \centering
    \includegraphics[width=0.65\linewidth]{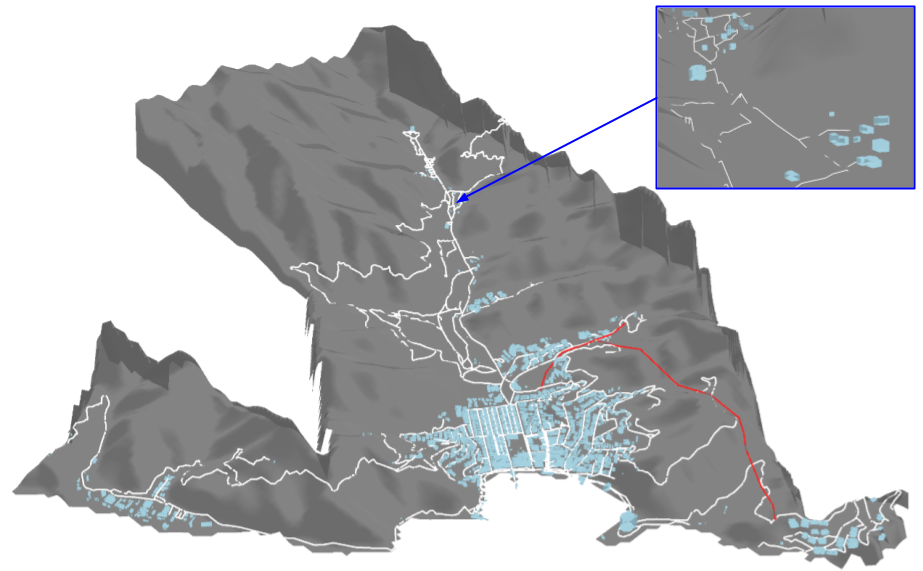}
    \caption{Visualized 3D model for the place of Avalon, Los Angeles County, United States. According to the log of our API, for this area, we collect 9,494 records of elevations, 929 road segments, 4 power lines, and 1,285 buildings with height. All the collected data is from open sources.}
    \label{fig:avalon_example_matlab}
\end{figure}

Fig.~\ref{fig:avalon_example_matlab} visualizes the 3D model for the place of Avalon, Los Angeles County, United States, when using MATLAB to request our API. We can see from the figure that Avalon is a mountainous area, with its buildings and roads mainly distributed in three districts. The largest district lies at the bottom of the valley, while the roads stretch up to the mountains. Though we can see that there are power lines connecting from the right corner to the middle part of Avalon, the power lines connecting the left corner in the figure are missing. This indicate that the power grid information from open sources - more specifically, from Open Street Map in our case - is not complete. It also suggests that the resulted 3D model for the same place can change because of updated information over the time in those open sources.

\section{Interoperability with other libraries}\label{sec:interoperability}

Within the context of the developed software suite, interoperability denotes the capacity of diverse systems to communicate and exchange data seamlessly. The \texttt{Cenergy3} Python library and its associated API achieve high levels of compatibility through a designed architecture centered on the following technical standards.

\subsection{Standardized coordinate systems}
The Python library and the deployed API adopt the \texttt{Web Mercator (EPSG:3857)} coordinate reference system as the universal standard for all 3D model generation. As the dominant projection for modern web mapping platforms - including Google Maps and OpenStreetMap - this choice ensures that the generated 3D coordinates are readily compatible with the majority of web-based visualization frameworks and geospatial platforms. Consequently, users can integrate the output into existing workflows without the extra computation or re-projections.

\subsection{Open data formats}
By anchoring and processing pipeline in open data sources and utilizing standard file formats for intermediate data states, the library and API ensure that its internal logic remains transparent and verifiable. The selection of \texttt{Overture Maps} and \texttt{OpenStreetMap} as primary data providers leverages a non-proprietary, community-driven data ecosystem. This approach ensures that the underlying data is subject to continuous peer scrutiny and remains accessible to the broad research community without restrictive licensing..

\subsection{API output and system integration}
The primary output of boh the lirary and the API is a \texttt{JSON}-serialized \texttt{Plotly} figure object. Given that JSON serves as the standard protocol for web-based data exchange, the generated models are highly versatile and can be used in the following environments: (i) web frontends. Utilizing \texttt{Plotly.js}, the model can be embedded into JavaScript-based web applications, providing interactive 3D visualization directly in the browser. (ii) data science environments. The \texttt{JSON} output can be instantly deserialized and manipulated within Python environments like Jupyter Notebooks, facilitating further statistical analysis or machine learning tasks, and (iii) cloud services. The API is built using the \texttt{FastAPI} framework, which simplifies deployment on containerized or serverless platforms. This results in a REstful API and allows the 3D model generation process to be triggered via standard \texttt{HTTP} requests. These design considerations ensure that the \texttt{Cenergy3} suite functions as a modular, ``plug-and-play'' component for any architectural system requiring on-demand, geospatially-aware 3D models.

\section{Conclusions}\label{sec:conclusion}

We presented an open software package for city energy 3D modeling. Our software retrieves, compiles, and integrates data from open sources, and provides the visualization of digital 3D urban energy model. The resulted model contains terrain, building semantics and heights, road networks, and power grids for an area of interest. We openly released our software package as an Python library and a free cloud-based API. We also developed the graphic user interface for our API. In this way, users of our software can access 3D models regardless of their programming skills and the computational power of their coding environments. We highlight that all the information and data collected and presented by our software is from open sources. This ensures that the outcomes of our software is generated on a basis scrutinized by a wide community, and we through this open software package also contribute to the open-source ecosystem. The source code of our software is publicly accessible via our Github page, and we envision more customized development for tailored purposes, \textit{e.g.}, in power grid management and monitoring, demand-supply analysis, detailed distribution power grid topology identification.

\bibliographystyle{unsrt}
\bibliography{references}

@ARTICLE{11364211,
  author={Zhang, Shiliang and Maharjan, Sabita and Strunz, Kai and Bryne, Jan Christian},
  journal={IEEE Data Descriptions}, 
  title={Descriptor: Norwegian Electricity in Geographic Dataset (NoreGeo)}, 
  year={2026},
  volume={3},
  number={},
  pages={82-92},
  doi={10.1109/IEEEDATA.2026.3658039}}

@article{zhang2025data,
  title={Data sharing, privacy and security considerations in the energy sector: A review from technical landscape to regulatory specifications},
  author={Zhang, Shiliang and Maharjan, Sabita and Bygrave, Lee Andrew and Yu, Shui},
  journal={arXiv preprint arXiv:2503.03539},
  year={2025}
}

@ARTICLE{9119753,
  author={Vargas-Munoz, John E. and Srivastava, Shivangi and Tuia, Devis and Falcão, Alexandre X.},
  journal={IEEE Geoscience and Remote Sensing Magazine}, 
  title={OpenStreetMap: Challenges and Opportunities in Machine Learning and Remote Sensing}, 
  year={2021},
  volume={9},
  number={1},
  pages={184-199},
  doi={10.1109/MGRS.2020.2994107}
}

@ARTICLE{oskar2025exploring,
  author={Oskar Våle and Shiliang Zhang and Sabita Maharjan and Gro Klæboe},
  journal={Artificial Intelligence Science and Engineering}, 
  title={Exploring the Interpretability of Forecasting Models for Energy Balancing Market}, 
  year={2025},
  volume={},
  number={},
  pages={}
}

@article{SALVALAI2024114500,
title = {From building energy modeling to urban building energy modeling: A review of recent research trend and simulation tools},
journal = {Energy and Buildings},
volume = {319},
pages = {114500},
year = {2024},
issn = {0378-7788},
doi = {https://doi.org/10.1016/j.enbuild.2024.114500},
author = {Graziano Salvalai and Yunxi Zhu and Marta {Maria Sesana}}
}

@article{el2024comprehensive,
  title={Comprehensive analysis of digital twins in smart cities: A 4200-paper bibliometric study},
  author={El-Agamy, Rasha F and Sayed, Hanaa A and AL Akhatatneh, Arwa M and Aljohani, Mansourah and Elhosseini, Mostafa},
  journal={Artificial Intelligence Review},
  volume={57},
  number={6},
  pages={154},
  year={2024},
  publisher={Springer}
}

@ARTICLE{10538433,
  author={Xu, Ruipeng and Zhang, Cuo and Zhang, Daming and Yang Dong, Zhao and Yip, Christine},
  journal={IEEE Transactions on Smart Grid}, 
  title={Adaptive Robust Load Restoration via Coordinating Distribution Network Reconfiguration and Mobile Energy Storage}, 
  year={2024},
  volume={15},
  number={6},
  pages={5485-5499},
  doi={10.1109/TSG.2024.3404776}
}

@article{hersbach2019era5,
  title={ERA5 hourly data on single levels from 1979 to present},
  author={Hersbach, Hans and Bell, Bill and Berrisford, Paul and Biavati, Gionata and Hor{\'a}nyi, And{\'a}s and Mu{\~n}oz Sabater, J and Nicolas, Julien and Peubey, Carole and Radu, Raluca and Rozum, Iryna and others},
  journal={Copernicus Climate Change Service (C3S) Climate Data Store (CDS)},
  year={2019}
}

@article{weinand2019spatial,
  title={Spatial high-resolution socio-energetic data for municipal energy system analyses},
  author={Weinand, Jann M and McKenna, Russell and Mainzer, Kai},
  journal={Scientific data},
  volume={6},
  number={1},
  pages={243},
  year={2019},
  publisher={Nature Publishing Group UK London}
}

@article{wylde2022cybersecurity,
  title={Cybersecurity, data privacy and blockchain: A review},
  author={Wylde, Vinden and Rawindaran, Nisha and Lawrence, John and Balasubramanian, Rushil and Prakash, Edmond and Jayal, Ambikesh and Khan, Imtiaz and Hewage, Chaminda and Platts, Jon},
  journal={SN computer science},
  volume={3},
  number={2},
  pages={127},
  year={2022},
  publisher={Springer}
}

@book{bennett2010openstreetmap,
  title={Open Street Map},
  author={Bennett, Jonathan},
  year={2010},
  publisher={Packt Publishing Ltd}
}

@article{boeing2025modeling,
  title={Modeling and analyzing urban networks and amenities with OSMnx},
  author={Boeing, Geoff},
  journal={Geographical Analysis},
  volume={57},
  number={4},
  pages={567--577},
  year={2025},
  publisher={Wiley Online Library}
}

@article{mukherjee2013evaluation,
  title={Evaluation of vertical accuracy of open source Digital Elevation Model (DEM)},
  author={Mukherjee, Sandip and Joshi, Pawan Kumar and Mukherjee, Samadrita and Ghosh, Aniruddha and Garg, RD and Mukhopadhyay, Anirban},
  journal={International Journal of Applied Earth Observation and Geoinformation},
  volume={21},
  pages={205--217},
  year={2013},
  publisher={Elsevier}
}

@inproceedings{krishnan2011opentopography,
  title={OpenTopography: a services oriented architecture for community access to LIDAR topography},
  author={Krishnan, Sriram and Crosby, Christopher and Nandigam, Viswanath and Phan, Minh and Cowart, Charles and Baru, Chaitanya and Arrowsmith, Ramon},
  booktitle={Proceedings of the 2nd international conference on computing for Geospatial Research \& Applications},
  pages={1--8},
  year={2011}
}

@incollection{kumar2023referencing,
  title={Referencing and coordinate systems in GIS},
  author={Kumar, Manish and Singh, RB and Singh, Anju and Pravesh, Ram and Majid, Syed Irtiza and Tiwari, Akash},
  booktitle={Geographic information systems in urban planning and management},
  pages={25--46},
  year={2023},
  publisher={Springer}
}

@article{stefanakis2017web,
  title={Web mercator and raster tile maps: two cornerstones of online map service providers},
  author={Stefanakis, Emmanuel},
  journal={Geomatica},
  volume={71},
  number={2},
  pages={100--109},
  year={2017},
  publisher={Elsevier}
}

@incollection{ujang2025introduction,
  title={Introduction to Geospatial Innovation for Smart City Development, Eco-synergy, and Urban Resurgence},
  author={Ujang, Uznir and Yadava, Ram Narayan},
  booktitle={Geospatial Innovation: Igniting Smart Cities, Eco-Synergy, and Urban Resurgence: Geospatial Technologies for Smart Cities},
  pages={3--15},
  year={2025},
  publisher={Springer}
}

@article{mallala2025forecasting,
  title={Forecasting global sustainable energy from renewable sources using random forest algorithm},
  author={Mallala, Balasubbareddy and Ahmed, Azka Ihtesham Uddin and Pamidi, Sastry V and Faruque, Md Omar and Reddy, Rajasekhar},
  journal={Results in Engineering},
  volume={25},
  pages={103789},
  year={2025},
  publisher={Elsevier}
}

@inproceedings{andres2025hot,
  title={Hot Spot Analysis: An Effective GIS Methodology and Energy Justice Tool to Visualize Power Outage in Vulnerable Communities},
  author={Andres, LaRico and Bui, Van-Hai and Koumpias, Antonios M and Su, Wencong},
  booktitle={2025 IEEE Power \& Energy Society General Meeting (PESGM)},
  pages={1--5},
  year={2025},
  organization={IEEE}
}

@article{ahshan2025geospatial,
  title={Geospatial mapping of large-scale electric power grids: A residual graph convolutional network-based approach with attention mechanism},
  author={Ahshan, Razzaqul and Abid, Md Shadman and Al-Abri, Mohammed},
  journal={Energy and AI},
  volume={20},
  pages={100486},
  year={2025},
  publisher={Elsevier}
}

@article{cheikh2026energy,
  title={Energy, scalability, data and security in massive IoT: Current landscape and future directions},
  author={Cheikh, Imane and Roy, S{\'e}bastien and Sabir, Essaid and Aouami, Rachid},
  journal={IEEE Internet of Things Journal},
  year={2026},
  publisher={IEEE}
}

@article{moroz2026urban,
  title={Urban morphology as a proxy for housing and infrastructure inequality: A machine learning approach using open building footprint data},
  author={Moroz, Cassiano Bastos and Thieken, Annegret H},
  journal={Computers, Environment and Urban Systems},
  volume={126},
  pages={102402},
  year={2026},
  publisher={Elsevier}
}

@article{khatiwada2026urban,
  title={Urban-Scale Feasibility Assessment of Green Roofs on Existing Buildings Using Remote Sensing, AI, and Non-Destructive Testing},
  author={Khatiwada, Prashidha and Sofi, Massoud and Zhou, Zhiyuan and Lumantarna, Elisa},
  journal={Sustainable Cities and Society: Advances},
  pages={100033},
  year={2026},
  publisher={Elsevier}
}

@article{quinones2025data,
  title={Data-driven design in the design process: A systematic literature review on challenges and opportunities},
  author={Qui{\~n}ones-G{\'o}mez, Juan Carlos and Mor, Enric and Chac{\'o}n, Jonathan},
  journal={International Journal of Human--Computer Interaction},
  volume={41},
  number={4},
  pages={2227--2252},
  year={2025},
  publisher={Taylor \& Francis}
}

@article{yan2023multi,
  title={Multi-label image recognition for electric power equipment inspection based on multi-scale dynamic graph convolution network},
  author={Yan, Yunfeng and Han, Yadong and Qi, Donglian and Lin, Jiajun and Yang, Zhi and Jin, Lingfeng},
  journal={Energy Reports},
  volume={9},
  pages={1928--1937},
  year={2023},
  publisher={Elsevier}
}

@article{kapp2023predicting,
  title={Predicting industrial building energy consumption with statistical and machine-learning models informed by physical system parameters},
  author={Kapp, Sean and Choi, Jun-Ki and Hong, Taehoon},
  journal={Renewable and Sustainable Energy Reviews},
  volume={172},
  pages={113045},
  year={2023},
  publisher={Elsevier}
}

@article{shiell2025overture,
  title={Overture: an open-source genomics data platform},
  author={Shiell, Mitchell and Bajari, Rosi and Andric, Dusan and Eubank, Jon and Chan, Brandon F and Richardsson, Anders J and Ali, Azher and Allabadi, Bashar and Alturmessov, Yelizar and Baker, Jared and others},
  journal={GigaScience},
  volume={14},
  pages={giaf038},
  year={2025},
  publisher={Oxford University Press}
}

@article{lathkar2023high,
  title={High-Performance Web Apps with FastAPI},
  author={Lathkar, Malhar},
  journal={California: Apress Berkeley},
  year={2023},
  publisher={Springer}
}

@book{biehl2016restful,
  title={RESTful API Design},
  author={Biehl, Matthias},
  volume={3},
  year={2016},
  publisher={API-University Press}
}
\end{document}